\documentclass[aip,jap,reprint]{revtex4-1}

\usepackage{amsmath,amssymb,booktabs,multirow,graphicx}

\begin{document}


\title{Compact electric-LC resonators for metamaterials} 

\author{Withawat Withayachumnankul}
\email[]{withawat@eleceng.adelaide.edu.au}
\altaffiliation{School of Electronic Engineering, Faculty of Engineering, King Mongkut's Institute of Technology Ladkrabang, Bangkok 10520, Thailand}

\author{Christophe Fumeaux}

\author{Derek Abbott}

\affiliation{School of Electrical \& Electronic Engineering, University of Adelaide, SA 5005, Australia}

\date{\today}

\begin{abstract}
Alternative designs to an electric-\textit{LC} (ELC) resonator, which is a type of metamaterial inclusion, are presented in this article. Fitting the resonator with an interdigital capacitor (IDC) helps to increase the total capacitance of the structure. In effect, its resonance frequency is shifted downwards. This implies a decreased overall resonator size with respect to its operating wavelength. As a result, the metamaterial, composed of an array of IDC-loaded ELC resonators with their collective electromagnetic response, possesses improved homogeneity and hence is less influenced by diffraction effects of individual cells. The impact of incorporating an IDC into ELC resonators in terms of the electrical size at resonance and other relevant properties are investigated through both simulation and experiment.
\end{abstract}


\maketitle 

\section{Introduction}

An electromagnetic metamaterial is a man-made composite material comprising a periodic array of subwavelength inclusions. Typically, a single metallic metamaterial inclusion can be considered as an \textit{LC} resonance circuit with its inductance and capacitance influenced by its shape and dimensions. These resonators can collectively exhibit macroscopically observed effective values of permittivity and/or permeability that are not found in natural materials. Various forms of resonant inclusions have been introduced to date, e.g., a split-ring resonators (SRR) for a magnetic response \cite{Pen99} or a pair of cut wires for negative refractive index \cite{Sha05}. Because of the possibility of engineering electromagnetic material properties, metamaterials offer immense opportunities in improving existing optical designs along with exploring unprecedented devices such as superlenses \cite{Pen00,Fan05} and invisibility cloaks \cite{Sch06b,Zha08}.

For these devices to function properly, the underlying metamaterials must be operated in the effective-medium regime, i.e., the lattice constant or unit cell size should be much smaller than $\lambda_0/4$, where $\lambda_0$ is the operating, i.e. resonant, wavelength. Under this condition a collection of metamaterial elements appears nearly homogeneous to incident waves and can be characterized by an effective permittivity and permeability. As the unit cell size approaches $\lambda_0/4$, diffraction effects and poor refraction become significant \cite{Cal05}. These parasitic effects are detrimental to the performance of metamaterials in quasi-optical applications. For example, a large unit cell size imposes a limitation on the subwavelength resolution of a metamaterial superlens \cite{Smi03}. When the cell size is comparable to or larger than a quarter-wavelength, the effective material parameters lose their relevance \cite{Lap07}. There are however limits to size reduction, because simply shrinking the volume or area of metamaterial inclusions reduces the capacitance and inductance and hence disturbs other important characteristics of metamaterials. In order to preserve these characteristics, the inductance and capacitance per unit area must be increased accordingly.

In this paper, a practical approach to reducing the electrical footprint of ELC resonators is presented. Section~\ref{sec:IE_review} reviews some existing solutions to reduce the size of metamaterial resonators. Section~\ref{sec:IE_idc_elc} describes the design of IDC-loaded ELC resonators, along with a mathematical analysis based on lumped element theory. In Section~\ref{sec:IE_experiment}, simulation and experimental results obtained from the proposed resonators are discussed in terms of the transmission characteristics and the effective medium properties. 

\vspace{-1mm}
\section{Existing solutions}\label{sec:IE_review}

Several approaches can be implemented to increase the ratio between the operating wavelength and the unit cell size, i.e. the effective medium ratio, for various types of metamaterial resonators. In common, these approaches are based on the idea of raising the overall inductance and capacitance, which are related to the resonance frequency through $f_0=1/(2\pi\sqrt{LC})$. A straightforward way is to change the feature sizes of the structure, i.e., shorten gaps or lengthen wires to increase the capacitance and inductance, respectively. However, the realization of this simple approach can be limited by fabrication tolerances. As an alternative, increasing the permittivity or permeability of the host dielectric leads to a larger effective capacitance or inductance \cite{Cal05}. Despite a substantial reduction in the resonance frequency, the availability of the host medium with a high permittivity or permeability and low loss is an important issue. Furthermore, at some point the effective permeability can no longer be increased, irrespective of the permeability of the substrate \cite{Ria09}. By integrating a surface-mounted capacitor onto a resonator, its resonance frequency can be tuned down significantly \cite{Ayd05,Ayd07}. However, loading lumped capacitive elements complicates the fabrication process and is typically limited by their physical size in the microwave regime.

A more sophisticated and specific approach involves redesigning the resonator pattern to accommodate a higher capacitance or inductance. As for example, a multiple SRR (MSRR) is an extension to a conventional SRR, within which smaller split rings are nested to increase the parallel capacitance \cite{Bil07}. With comparable dimensions, a spiral resonator (SR) has a larger capacitance than does a typical SRR by at least fourfold \cite{Bae04}. Essentially, an increment in the structural capacitance results in a realizable unit cell size down to $\lambda_0$/40 for an MSRR and $\lambda_0$/250 for an SR \cite{Bil07}. Fractal-based metamaterials with magnetic-field coupling similar to double SRR's have been reported \cite{Len09}. The increased perimeter of the structure as a result of the fractal self-similarity leads to larger inductance and capacitance and a reduction in the resonance frequency. S-ring resonators that provide a double-negative response can be shrunken from $\lambda_0/6$ to $\lambda_0/15$ by winding parallel strips to increase the capacitance \cite{Che06}. It is interesting to note that all the redesigned structures do not provide a pure electric resonance.

\vspace{-1mm}
\section{IDC-loaded ELC resonators}\label{sec:IE_idc_elc}

\begin{figure}
\includegraphics{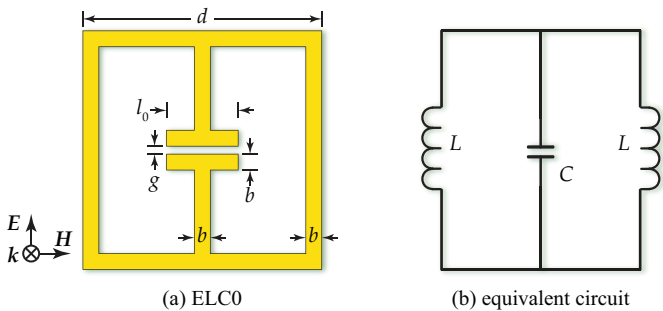}
\caption{Electric-\textit{LC} resonator. (a) A typical ELC resonator is composed of a center capacitive gap connected to two inductive loops. (b) An equivalent circuit of the resonator constitutes an \textit{LC} resonator (the resistance is neglected here) \cite{Sch06}.}
\label{fig:IE_elc0}
\end{figure}

As a counterpart of the conventional magnetic SRR, an ELC resonator as shown in Fig.~\ref{fig:IE_elc0}(a) provides a pure electric resonance with neither magnetic nor magnetoelectric responses, since the collective magnetic flux is nullified by virtue of the resonator's mirror symmetry \cite{Sch06,Pad07}. In the quasistatic limit, where the resonator size is much smaller than its operating wavelength, an ELC resonator can be approximated by the inductance and capacitance in the form of an \textit{LC} resonance circuit, as illustrated in  Fig.~\ref{fig:IE_elc0}(b). During operation, an incident electric field with polarization perpendicular to the gap excites the capacitor to yield an electric resonance at $f_0=1/(\pi\sqrt{2LC})$. It was suggested that the resonance frequency of an ELC resonator can be tuned down by introducing more loops, or equivalently more inductance, to both arms \cite{Sch06}. However, the method requires additional fabrication steps, as the suggested structure is multilayered. Other general approaches reviewed in the previous section might be adopted, with their respective limitations, to reduce the electrical size of the resonator.

\begin{figure}
\includegraphics{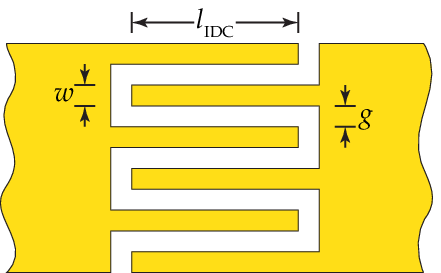}
\caption{Interdigital capacitor (IDC) with 6 fingers. The finger length $l_{\rm IDC}$ extends over the overlapping portion of all fingers.}
\label{fig:IE_IDC}
\end{figure}

A practical redesign of the structure presented in this article involves replacing the normal capacitive gap in the center of an ELC resonator with an interdigital capacitor (IDC), illustrated in Fig.~\ref{fig:IE_IDC}, to increase the capacitance. Essentially, IDC's are widely used as lumped elements in microwave circuits \cite{Gup96} with an aim to decrease their circuit board footprint. As such, its concept well suits metamaterials, where the unit cell needs to be much smaller than the operating wavelength. Fabrication of IDC's can be readily carried out with standard photolithographic techniques, since the feature size of an IDC does not need to be finer than the gap width of an ordinary ELC resonator to enhance the capacitance. The single-layered design of IDC-loaded ELC resonators eliminates additional fabrication steps that might be required in other approaches.

The idea of using IDC's with metamaterials has been realized in the transmission-line approach. Composite right/left handed transmission lines (CRLH-TL's), which exhibit a band of negative phase velocity, are typically fitted with a set of IDC's that act as series capacitors \cite{Cal04,San04}. In addition, IDC's are also incorporated into microstrip-coupled tunable SRRs aimed at enhancing the local electric field strength at the capacitors' gaps \cite{Hou10}. 

\begin{figure}
\centering\includegraphics{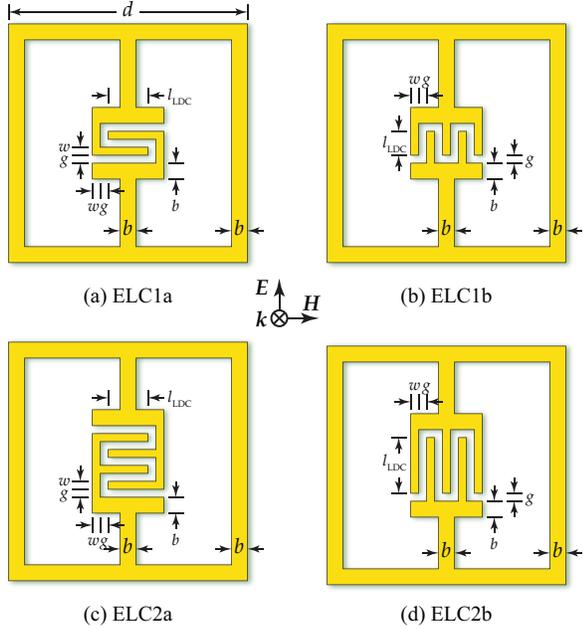}
\caption{Two variants of IDC-loaded ELC resonators. The IDC's gaps align (a,c) in perpendicular to and (b,d) in parallel with the direction of an incident electric field.}
\label{fig:IE_elcs}
\end{figure}

Two variants for IDC-loaded ELC resonators are proposed and studied in this article. For the first configuration shown in Fig.~\ref{fig:IE_elcs}(a,c), the capacitor's gaps are oriented perpendicularly to the intended polarization of operation, and the two ports are positioned at the outermost fingers on both sides. Another configuration in Fig.~\ref{fig:IE_elcs}(b,d) has the gaps oriented in parallel to the polarization. The ports are located at the terminal strips (the strips that join fingers together). For the first configuration, the local electric field inside the capacitive gap aligns with the electric polarization. Hence, this configuration is expected to perform better in terms of cell size reduction. It is worth noting that the mirror symmetry of the resonator is broken in the first configuration. However, this does not affect the cancellation of the magnetic dipoles, since the area of the two current loops of the resonator are still identical.

Analytically, the capacitance of the IDC shown in Fig.~\ref{fig:IE_IDC} is a function of the finger length $l_{\rm IDC}$, total number of fingers $N$, line width $w$, gap width $g$, and effective dielectric constant $\epsilon_{\rm re}$, as indicated in the following formula \cite{Gup96,Bah03}:
\begin{eqnarray}\label{eq:IE_cap_fn}
C_{\rm IDC}&=&\frac{\epsilon_{\rm re}10^{-3}}{18\pi}\frac{K(k)}{K'(k)}(N-1)l_{\rm IDC}\quad(\mathrm{pF})\;,
\end{eqnarray}
where the approximated ratio between the elliptic integrals of first kind $K(k)$ and its complement $K'(k)$ reads
\begin{eqnarray}
\frac{K(k)}{K'(k)}&=&\left\{ \begin{array}{ll}
\frac{1}{\pi}\ln\left[2\frac{1+\sqrt{k}}{1-\sqrt{k}}\right] & \mathrm{for}\quad 0.707\leq k \leq 1\\
\frac{\pi}{\ln\left[2\frac{1+\sqrt{k'}}{1-\sqrt{k'}}\right]} & \mathrm{for}\quad 0\leq k < 0.707\;,\\
\end{array} \right.
\end{eqnarray}
and $k'=\sqrt{1-k^2}$; $k=\tan^2\left[0.25w\pi/(w+g)\right]$. All the lengths are in microns. An ordinary parallel-strip capacitor can be approximated as an IDC with $N=2$, the capacitance of which can be deduced from Eq.~(\ref{eq:IE_cap_fn}) as
\begin{eqnarray}
C_0&=&\frac{\epsilon_{\rm re}10^{-3}}{18\pi}\frac{K(k)}{K'(k)}l_0\quad(\mathrm{pF})\;,
\end{eqnarray}
where $l_0$ is the strip length (see Fig.~\ref{fig:IE_elc0}). Provided that an IDC and a parallel-strip capacitor possess the same line width, gap width, and substrate type, their capacitances can be related through
\begin{eqnarray}\label{eq:IE_cap_relation}
C_{\rm IDC}&=&(N-1)\frac{l_{\rm IDC}}{l_0}C_0\;.
\end{eqnarray}
The factor $l_{\rm IDC}/l_0$ compensates the difference between the finger length and strip length. 

For an ELC resonator, if the inductance loop remains unchanged, it can be estimated from Eq.~(\ref{eq:IE_cap_relation}) that the new resonance frequency $f_{0,{\rm new}}$ after IDC loading equals
\begin{eqnarray}\label{eq:IE_freq_relation}
f_{0,{\rm new}}&=&\sqrt{\frac{l_0}{l_{\rm IDC}(N-1)}}f_0\;,
\end{eqnarray}
where $f_0$ is the resonance frequency of a conventional ELC resonator. This simple model gives an impression for the expected change in the resonance frequency of IDC-loaded ELC resonators.

\vspace{-1mm}
\section{Results}\label{sec:IE_experiment}

\subsection{Transmission characteristics}

\begin{table}
	\caption{The structural parameters and resonance frequency of the samples under test.}
	\label{tab:IE_resonance}
	\vspace{+.5cm}
	\centering
  \begin{tabular*}{\columnwidth}
     {@{\extracolsep{\fill}}lcccccc}
\hline \multirow{2}{*}{Sample}& 
\multirow{2}{*}{$N$} &\multirow{2}{*}{$l_{\rm IDC}$ (mm)} & \multicolumn{3}{c}{$f_0$ (GHz)} & \multirow{2}{*}{$\lambda_0/a$}
\\\cline{4-6}
&&& simulated & estimated & measured \\
\hline
ELC0 & 2 & $l_0$ = 3.6 & 3.16 & - & 3.14 & 6.8 \\
ELC1a & 4 & 2.0 & 2.51 & 2.45 & 2.48 & 8.7\\
ELC1b & 5 & 1.2 & 2.67 & 2.74 & 2.64 & 8.1 \\
ELC2a & 6 & 2.0 & 2.11 & 1.90 & 2.10 & 10.2\\
ELC2b & 5 & 2.8 & 2.19 & 1.79 & 2.24 & 9.6\\
 \hline

\end{tabular*}
\end{table}

\begin{figure}[b]
\includegraphics{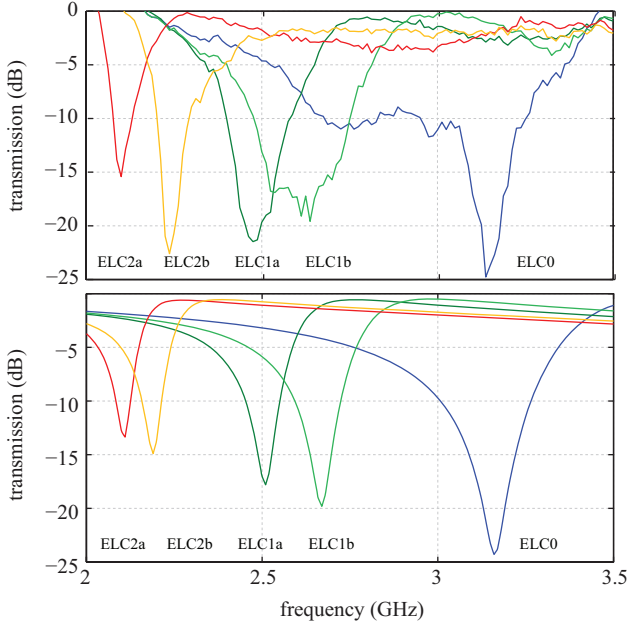}
\caption{Transmission profiles of the resonators from (top) the experiment and (bottom) the simulation. The two graphs share the horizontal scale.}
\label{fig:IE_transmission}
\end{figure}

The four designs of IDC-loaded ELC resonators in Fig.~\ref{fig:IE_elcs}, along with a corresponding conventional resonator in Fig.~\ref{fig:IE_elc0}, are fabricated and characterized in the microwave regime. For the sake of comparison, the five designs share the unit cell size $a$, gap width $g$, and cross-sectional length of the loaded capacitor. It is suggested that the finger width should be equal to the gap width, or $w=g$, to maximize the capacitance density \cite{All70}. In details, the structural parameters common to all designs are $w=g=0.4$~mm, $b=0.8$~mm, $d=12$~mm, and $a=14$~mm. Other parameters specific to each design are given in Table~\ref{tab:IE_resonance}. Each planar metamaterial comprises an array of 9$\times$9 identical resonators, made of copper with a thickness of 35.6~$\mu$m (1.4~mil). The planar substrate is an FR4 epoxy board with a thickness of 0.8~mm, a measured dielectric constant of 4.2, and a reported loss tangent of 0.02. In the experiment, a metamaterial sample is located between two horn antennas facing each other, from which the transmission through the sample is measured and compared to free-space transmission. The measurement results are given in Fig~\ref{fig:IE_transmission}(top).

In order to verify the experimental results, the simulation for IDC-loaded ELC resonators is performed with a finite-element-based electromagnetic solver, Ansoft HFSS. Periodic boundary conditions are utilized for the transverse boundaries to replicate an infinite planar array of the resonators. Two ports at the open ends allows to determine the response of the sample to a plane wave incident normally to the array. As shown in Fig.~\ref{fig:IE_transmission} the simulation results are in general agreement with the experimental data. The discrepancies are attributed to the finite size of the fabricated array and the nonuniformity among the resonators due to fabrication tolerances. Table~\ref{tab:IE_resonance} indicates a general agreement in the resonance frequencies obtained from the simulation and the experiment. In addition, Eq.~\ref{eq:IE_freq_relation} can provide a rough estimation in the resonance frequency. Note that in a 3D configuration, the interaction between layers might shift the resonance slightly.

It is clear from Fig.~\ref{fig:IE_transmission} and Table~\ref{tab:IE_resonance} that the resonance frequency of IDC-loaded ELC resonators is remarkably lower than that of the original design. As a consequence, the coupling strength reduces for those structures with a higher capacitance \cite{Sch06,Ayd07}. With a comparable capacitance area and density, sample ELC1a (ELC2a) has a lower resonance frequency compared to sample ELC1b (ELC2b). A reduction in resonance frequency is stronger when the gaps of the IDC are perpendicular to the polarization, as in ELC1a and ELC2a, which can be attributed to a better coupling between the incident electric field and the field in the capacitors. In terms of the effective medium ratio, the maximum improvement can be observed in sample ELC2a, for which $\lambda_0/a$ equals 10.2, in comparison to 6.8 of the conventional ELC design (ELC0). The effective medium ratio of other samples is listed in Table~\ref{tab:IE_resonance}. Note that the original ELC design \cite{Sch06} has a unit cell size of $\lambda_0/5.7$.

\subsection{Effective medium properties}

\begin{figure}
\includegraphics{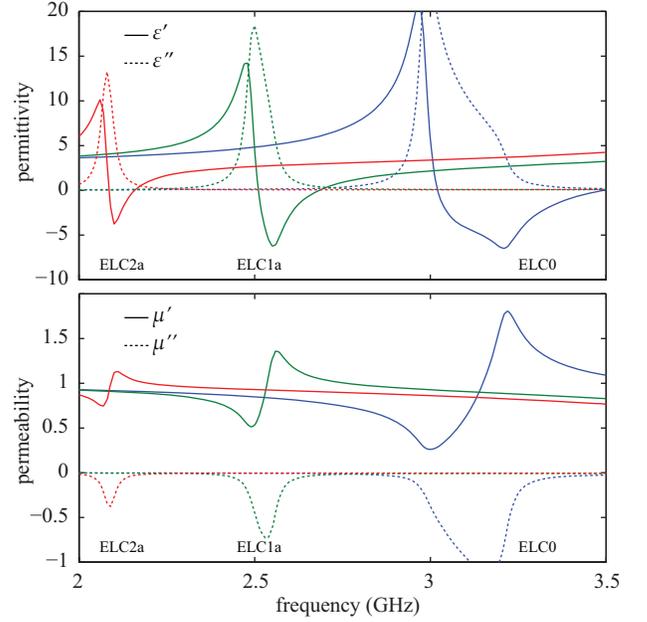}
\caption{The simulated effective medium properties of the samples; (top) the effective permittivity, and (bottom) the effective permeability. The two graphs share the horizontal scale.}
\label{fig:IE_param}
\end{figure}

In order to provide further insight, the samples are characterized for their effective medium parameters. It is worth nothing that the parameters are evaluated on the basis of planar metamaterials. Hence, these parameters would need to be fine-tuned for 3D operation to take into account the weak inter-layer coupling. Here, the effective permittivity and permeability are extracted from the simulated transmission/reflection magnitude and phase using the method of Chen~\textit{et al.} \cite{Che04}. For phase de-embedding, it is assumed that the thickness of a metamaterial sample in the direction of wave propagation is equal to the unit cell size or 14~mm in the present case \cite{Sch06}. Fig.~\ref{fig:IE_param} shows the effective permittivity and permeability of the selected resonators. It is obvious that the resonance frequency in the permittivity curve is lowered by the influence of IDC loading. The permittivity further illustrates a lower resonance strength and a higher effective loss tangent in IDC-loaded ELC resonators. Despite that, the negative value of the permittivity is appropriate for typical applications requiring $\epsilon=-1$. In addition, further simulation results (not shown here) indicate that with a lower-loss substrate, the effective loss tangent can be significantly reduced.

As discussed earlier, ELC resonators essentially possess no magnetic response, i.e., the real part of the permeability is close to unity over the frequencies of interest. However, the retrieved parameters do not strictly comply with this principle. As shown in Fig.~\ref{fig:IE_param}, the real permeability becomes anti-resonant, and the imaginary part is negative. In fact, these anomalies are artifacts introduced during parameter extraction due to the inhomogeneity of metamaterials \cite{Smi05,Sch06}. The reduction of these anomalies for ELC1a and ELC2a results from a lower dispersion and the higher homogeneity in these structures.

\vspace{-1mm}
\section{Conclusion}

This article proposes to miniaturize ELC resonators through IDC loading. In the experiment, a set of IDC-loaded ELC samples is fabricated and characterized in the microwave frequency range. The measurement data, in agreement with the simulation results, reveal a significant improvement in the effective medium ratio of these IDC-loaded resonators. The parameter retrieval suggests a tradeoff between the electrical size and absorption in the proposed structure. However, this issue can be partly alleviated by using a substrate with lower loss. An implication of metamaterial size reduction is the structural homogeneity, which leads to lower parasitic effects and hence a higher performance for quasi-optical applications.

The IDC-loading approach can be used in conjunction with other approaches to minimize the electrical size of ELC resonators. Apart from that, the proposed approach can be implemented when other options are not available due to fabrication limits. This is particularly relevant for terahertz metamaterials, which have a restricted range of fabrication techniques \cite{Wit09}. In addition, the IDC loading can be applied to other resonance-based metamaterials as well. Last but not least, the reduction of the resonator size is not only beneficial for metamaterial homogeneity but also useful in applications of metamaterial-based antennas and sensors, where a smaller electrical size of these structures is of a prime importance. 

\vspace{-0.3cm}
\begin{acknowledgments}
The authors acknowledge Pavel Simcik for his technical assistance. This research was supported under the Australian Research Council \textit{Discovery Projects} funding scheme (project number DP1095151).
\end{acknowledgments}


\begin{thebibliography}{27}%
\makeatletter
\providecommand \@ifxundefined [1]{%
 \@ifx{#1\undefined}
}%
\providecommand \@ifnum [1]{%
 \ifnum #1\expandafter \@firstoftwo
 \else \expandafter \@secondoftwo
 \fi
}%
\providecommand \@ifx [1]{%
 \ifx #1\expandafter \@firstoftwo
 \else \expandafter \@secondoftwo
 \fi
}%
\providecommand \natexlab [1]{#1}%
\providecommand \enquote  [1]{``#1''}%
\providecommand \bibnamefont  [1]{#1}%
\providecommand \bibfnamefont [1]{#1}%
\providecommand \citenamefont [1]{#1}%
\providecommand \href@noop [0]{\@secondoftwo}%
\providecommand \href [0]{\begingroup \@sanitize@url \@href}%
\providecommand \@href[1]{\@@startlink{#1}\@@href}%
\providecommand \@@href[1]{\endgroup#1\@@endlink}%
\providecommand \@sanitize@url [0]{\catcode `\\12\catcode `\$12\catcode
  `\&12\catcode `\#12\catcode `\^12\catcode `\_12\catcode `\%12\relax}%
\providecommand \@@startlink[1]{}%
\providecommand \@@endlink[0]{}%
\providecommand \url  [0]{\begingroup\@sanitize@url \@url }%
\providecommand \@url [1]{\endgroup\@href {#1}{\urlprefix }}%
\providecommand \urlprefix  [0]{URL }%
\providecommand \Eprint [0]{\href }%
\providecommand \doibase [0]{http://dx.doi.org/}%
\providecommand \selectlanguage [0]{\@gobble}%
\providecommand \bibinfo  [0]{\@secondoftwo}%
\providecommand \bibfield  [0]{\@secondoftwo}%
\providecommand \translation [1]{[#1]}%
\providecommand \BibitemOpen [0]{}%
\providecommand \bibitemStop [0]{}%
\providecommand \bibitemNoStop [0]{.\EOS\space}%
\providecommand \EOS [0]{\spacefactor3000\relax}%
\providecommand \BibitemShut  [1]{\csname bibitem#1\endcsname}%
\let\auto@bib@innerbib\@empty
\bibitem [{\citenamefont {Pendry}\ \emph {et~al.}(1999)\citenamefont {Pendry},
  \citenamefont {Holden}, \citenamefont {Robbins},\ and\ \citenamefont
  {Stewart}}]{Pen99}%
  \BibitemOpen
  \bibfield  {author} {\bibinfo {author} {\bibfnamefont {J.~B.}\ \bibnamefont
  {Pendry}}, \bibinfo {author} {\bibfnamefont {A.~J.}\ \bibnamefont {Holden}},
  \bibinfo {author} {\bibfnamefont {D.~J.}\ \bibnamefont {Robbins}}, \ and\
  \bibinfo {author} {\bibfnamefont {W.~J.}\ \bibnamefont {Stewart}},\
  }\href@noop {} {\bibfield  {journal} {\bibinfo  {journal} {IEEE Transactions
  on Microwave Theory and Techniques}\ }\textbf {\bibinfo {volume} {47}},\
  \bibinfo {pages} {2075} (\bibinfo {year} {1999})}\BibitemShut {NoStop}%
\bibitem [{\citenamefont {Shalaev}\ \emph {et~al.}(2005)\citenamefont
  {Shalaev}, \citenamefont {Cai}, \citenamefont {Chettiar}, \citenamefont
  {Yuan}, \citenamefont {Sarychev}, \citenamefont {Drachev},\ and\
  \citenamefont {Kildishev}}]{Sha05}%
  \BibitemOpen
  \bibfield  {author} {\bibinfo {author} {\bibfnamefont {V.}~\bibnamefont
  {Shalaev}}, \bibinfo {author} {\bibfnamefont {W.}~\bibnamefont {Cai}},
  \bibinfo {author} {\bibfnamefont {U.}~\bibnamefont {Chettiar}}, \bibinfo
  {author} {\bibfnamefont {H.}~\bibnamefont {Yuan}}, \bibinfo {author}
  {\bibfnamefont {A.}~\bibnamefont {Sarychev}}, \bibinfo {author}
  {\bibfnamefont {V.}~\bibnamefont {Drachev}}, \ and\ \bibinfo {author}
  {\bibfnamefont {A.}~\bibnamefont {Kildishev}},\ }\href@noop {} {\bibfield
  {journal} {\bibinfo  {journal} {Optics Letters}\ }\textbf {\bibinfo {volume}
  {30}},\ \bibinfo {pages} {3356} (\bibinfo {year} {2005})}\BibitemShut
  {NoStop}%
\bibitem [{\citenamefont {Pendry}(2000)}]{Pen00}%
  \BibitemOpen
  \bibfield  {author} {\bibinfo {author} {\bibfnamefont {J.~B.}\ \bibnamefont
  {Pendry}},\ }\href {\doibase 10.1103/PhysRevLett.85.3966} {\bibfield
  {journal} {\bibinfo  {journal} {Physical Review Letters}\ }\textbf {\bibinfo
  {volume} {85}},\ \bibinfo {pages} {3966} (\bibinfo {year}
  {2000})}\BibitemShut {NoStop}%
\bibitem [{\citenamefont {Fang}\ \emph {et~al.}(2005)\citenamefont {Fang},
  \citenamefont {Lee}, \citenamefont {Sun},\ and\ \citenamefont
  {Zhang}}]{Fan05}%
  \BibitemOpen
  \bibfield  {author} {\bibinfo {author} {\bibfnamefont {N.}~\bibnamefont
  {Fang}}, \bibinfo {author} {\bibfnamefont {H.}~\bibnamefont {Lee}}, \bibinfo
  {author} {\bibfnamefont {C.}~\bibnamefont {Sun}}, \ and\ \bibinfo {author}
  {\bibfnamefont {X.}~\bibnamefont {Zhang}},\ }\href@noop {} {\bibfield
  {journal} {\bibinfo  {journal} {Science}\ }\textbf {\bibinfo {volume}
  {308}},\ \bibinfo {pages} {534} (\bibinfo {year} {2005})}\BibitemShut
  {NoStop}%
\bibitem [{\citenamefont {Schurig}\ \emph {et~al.}(2006)\citenamefont
  {Schurig}, \citenamefont {Mock}, \citenamefont {Justice}, \citenamefont
  {Cummer}, \citenamefont {Pendry}, \citenamefont {Starr},\ and\ \citenamefont
  {Smith}}]{Sch06b}%
  \BibitemOpen
  \bibfield  {author} {\bibinfo {author} {\bibfnamefont {D.}~\bibnamefont
  {Schurig}}, \bibinfo {author} {\bibfnamefont {J.~J.}\ \bibnamefont {Mock}},
  \bibinfo {author} {\bibfnamefont {B.~J.}\ \bibnamefont {Justice}}, \bibinfo
  {author} {\bibfnamefont {S.~A.}\ \bibnamefont {Cummer}}, \bibinfo {author}
  {\bibfnamefont {J.~B.}\ \bibnamefont {Pendry}}, \bibinfo {author}
  {\bibfnamefont {A.~F.}\ \bibnamefont {Starr}}, \ and\ \bibinfo {author}
  {\bibfnamefont {D.~R.}\ \bibnamefont {Smith}},\ }\href@noop {} {\bibfield
  {journal} {\bibinfo  {journal} {Science}\ }\textbf {\bibinfo {volume}
  {314}},\ \bibinfo {pages} {977} (\bibinfo {year} {2006})}\BibitemShut
  {NoStop}%
\bibitem [{\citenamefont {Zharova}\ \emph {et~al.}(2008)\citenamefont
  {Zharova}, \citenamefont {Shadrivov}, \citenamefont {Zharov},\ and\
  \citenamefont {Kivshar}}]{Zha08}%
  \BibitemOpen
  \bibfield  {author} {\bibinfo {author} {\bibfnamefont {N.~A.}\ \bibnamefont
  {Zharova}}, \bibinfo {author} {\bibfnamefont {I.~V.}\ \bibnamefont
  {Shadrivov}}, \bibinfo {author} {\bibfnamefont {A.~A.}\ \bibnamefont
  {Zharov}}, \ and\ \bibinfo {author} {\bibfnamefont {Y.}~\bibnamefont
  {Kivshar}},\ }\href@noop {} {\bibfield  {journal} {\bibinfo  {journal}
  {Optics Express}\ }\textbf {\bibinfo {volume} {16}},\ \bibinfo {pages}
  {21369} (\bibinfo {year} {2008})}\BibitemShut {NoStop}%
\bibitem [{\citenamefont {Caloz}, \citenamefont {Lai},\ and\ \citenamefont
  {Itoh}(2005)}]{Cal05}%
  \BibitemOpen
  \bibfield  {author} {\bibinfo {author} {\bibfnamefont {C.}~\bibnamefont
  {Caloz}}, \bibinfo {author} {\bibfnamefont {A.}~\bibnamefont {Lai}}, \ and\
  \bibinfo {author} {\bibfnamefont {T.}~\bibnamefont {Itoh}},\ }\href@noop {}
  {\bibfield  {journal} {\bibinfo  {journal} {New Journal of Physics}\ }\textbf
  {\bibinfo {volume} {7}},\ \bibinfo {pages} {167} (\bibinfo {year}
  {2005})}\BibitemShut {NoStop}%
\bibitem [{\citenamefont {Smith}\ \emph {et~al.}(2003)\citenamefont {Smith},
  \citenamefont {Schurig}, \citenamefont {Rosenbluth}, \citenamefont {Schultz},
  \citenamefont {Ramakrishna},\ and\ \citenamefont {Pendry}}]{Smi03}%
  \BibitemOpen
  \bibfield  {author} {\bibinfo {author} {\bibfnamefont {D.~R.}\ \bibnamefont
  {Smith}}, \bibinfo {author} {\bibfnamefont {D.}~\bibnamefont {Schurig}},
  \bibinfo {author} {\bibfnamefont {M.}~\bibnamefont {Rosenbluth}}, \bibinfo
  {author} {\bibfnamefont {S.}~\bibnamefont {Schultz}}, \bibinfo {author}
  {\bibfnamefont {S.~A.}\ \bibnamefont {Ramakrishna}}, \ and\ \bibinfo {author}
  {\bibfnamefont {J.~B.}\ \bibnamefont {Pendry}},\ }\href@noop {} {\bibfield
  {journal} {\bibinfo  {journal} {Applied Physics Letters}\ }\textbf {\bibinfo
  {volume} {82}},\ \bibinfo {pages} {1506} (\bibinfo {year}
  {2003})}\BibitemShut {NoStop}%
\bibitem [{\citenamefont {M.~Lapine}\ and\ \citenamefont
  {Tretyakov}(2007)}]{Lap07}%
  \BibitemOpen
  \bibfield  {author} {\bibinfo {author} {\bibfnamefont {M.}~\bibnamefont
  {M.~Lapine}}\ and\ \bibinfo {author} {\bibfnamefont {S.}~\bibnamefont
  {Tretyakov}},\ }\href@noop {} {\bibfield  {journal} {\bibinfo  {journal} {IET
  Microwaves, Antennas \& Propagation}\ }\textbf {\bibinfo {volume} {1}},\
  \bibinfo {pages} {3} (\bibinfo {year} {2007})}\BibitemShut {NoStop}%
\bibitem [{\citenamefont {Rialet}\ \emph {et~al.}(2009)\citenamefont {Rialet},
  \citenamefont {Sharaiha}, \citenamefont {Tarot},\ and\ \citenamefont
  {Delaveaud}}]{Ria09}%
  \BibitemOpen
  \bibfield  {author} {\bibinfo {author} {\bibfnamefont {D.}~\bibnamefont
  {Rialet}}, \bibinfo {author} {\bibfnamefont {A.}~\bibnamefont {Sharaiha}},
  \bibinfo {author} {\bibfnamefont {A.-C.}\ \bibnamefont {Tarot}}, \ and\
  \bibinfo {author} {\bibfnamefont {C.}~\bibnamefont {Delaveaud}},\ }in\
  \href@noop {} {\emph {\bibinfo {booktitle} {Third European Conference on
  Antennas and Propagation (EuCAP)}}}\ (\bibinfo {year} {2009})\ pp.\ \bibinfo
  {pages} {3163--3166}\BibitemShut {NoStop}%
\bibitem [{\citenamefont {Aydin}\ \emph {et~al.}(2005)\citenamefont {Aydin},
  \citenamefont {Bulu}, \citenamefont {Guven}, \citenamefont {Kafesaki},
  \citenamefont {Soukoulis},\ and\ \citenamefont {Ozbay}}]{Ayd05}%
  \BibitemOpen
  \bibfield  {author} {\bibinfo {author} {\bibfnamefont {K.}~\bibnamefont
  {Aydin}}, \bibinfo {author} {\bibfnamefont {I.}~\bibnamefont {Bulu}},
  \bibinfo {author} {\bibfnamefont {K.}~\bibnamefont {Guven}}, \bibinfo
  {author} {\bibfnamefont {M.}~\bibnamefont {Kafesaki}}, \bibinfo {author}
  {\bibfnamefont {C.~M.}\ \bibnamefont {Soukoulis}}, \ and\ \bibinfo {author}
  {\bibfnamefont {E.}~\bibnamefont {Ozbay}},\ }\href@noop {} {\bibfield
  {journal} {\bibinfo  {journal} {New Journal of Physics}\ }\textbf {\bibinfo
  {volume} {7}},\ \bibinfo {pages} {168} (\bibinfo {year} {2005})}\BibitemShut
  {NoStop}%
\bibitem [{\citenamefont {Aydin}\ and\ \citenamefont {Ozbay}(2007)}]{Ayd07}%
  \BibitemOpen
  \bibfield  {author} {\bibinfo {author} {\bibfnamefont {K.}~\bibnamefont
  {Aydin}}\ and\ \bibinfo {author} {\bibfnamefont {E.}~\bibnamefont {Ozbay}},\
  }\href@noop {} {\bibfield  {journal} {\bibinfo  {journal} {Journal of Applied
  Physics}\ }\textbf {\bibinfo {volume} {101}},\ \bibinfo {pages} {024911}
  (\bibinfo {year} {2007})}\BibitemShut {NoStop}%
\bibitem [{\citenamefont {Bilotti}, \citenamefont {Toscano},\ and\
  \citenamefont {Vegni}(2007)}]{Bil07}%
  \BibitemOpen
  \bibfield  {author} {\bibinfo {author} {\bibfnamefont {F.}~\bibnamefont
  {Bilotti}}, \bibinfo {author} {\bibfnamefont {A.}~\bibnamefont {Toscano}}, \
  and\ \bibinfo {author} {\bibfnamefont {L.}~\bibnamefont {Vegni}},\
  }\href@noop {} {\bibfield  {journal} {\bibinfo  {journal} {IEEE Transactions
  on Antennas and Propagation}\ }\textbf {\bibinfo {volume} {55}},\ \bibinfo
  {pages} {2258} (\bibinfo {year} {2007})}\BibitemShut {NoStop}%
\bibitem [{\citenamefont {Baena}, \citenamefont {Marqu\'es},\ and\
  \citenamefont {Medina}(2004)}]{Bae04}%
  \BibitemOpen
  \bibfield  {author} {\bibinfo {author} {\bibfnamefont {J.~D.}\ \bibnamefont
  {Baena}}, \bibinfo {author} {\bibfnamefont {R.}~\bibnamefont {Marqu\'es}}, \
  and\ \bibinfo {author} {\bibfnamefont {F.}~\bibnamefont {Medina}},\
  }\href@noop {} {\bibfield  {journal} {\bibinfo  {journal} {Physical Review
  B}\ }\textbf {\bibinfo {volume} {69}},\ \bibinfo {pages} {014402} (\bibinfo
  {year} {2004})}\BibitemShut {NoStop}%
\bibitem [{\citenamefont {Lenz}\ and\ \citenamefont {Henke}(2009)}]{Len09}%
  \BibitemOpen
  \bibfield  {author} {\bibinfo {author} {\bibfnamefont {E.}~\bibnamefont
  {Lenz}}\ and\ \bibinfo {author} {\bibfnamefont {H.}~\bibnamefont {Henke}},\
  }\href@noop {} {\bibfield  {journal} {\bibinfo  {journal} {Journal of Optics
  A: Pure and Applied Optics}\ }\textbf {\bibinfo {volume} {11}},\ \bibinfo
  {pages} {114021} (\bibinfo {year} {2009})}\BibitemShut {NoStop}%
\bibitem [{\citenamefont {Chen}\ \emph {et~al.}(2006)\citenamefont {Chen},
  \citenamefont {Ran}, \citenamefont {Wu}, \citenamefont {Kong},\ and\
  \citenamefont {Grzegorczyk}}]{Che06}%
  \BibitemOpen
  \bibfield  {author} {\bibinfo {author} {\bibfnamefont {H.}~\bibnamefont
  {Chen}}, \bibinfo {author} {\bibfnamefont {L.}~\bibnamefont {Ran}}, \bibinfo
  {author} {\bibfnamefont {B.-I.}\ \bibnamefont {Wu}}, \bibinfo {author}
  {\bibfnamefont {J.~A.}\ \bibnamefont {Kong}}, \ and\ \bibinfo {author}
  {\bibfnamefont {T.~M.}\ \bibnamefont {Grzegorczyk}},\ }\href@noop {}
  {\bibfield  {journal} {\bibinfo  {journal} {Progress in Electromagnetics
  Research}\ }\textbf {\bibinfo {volume} {66}},\ \bibinfo {pages} {179}
  (\bibinfo {year} {2006})}\BibitemShut {NoStop}%
\bibitem [{\citenamefont {Schurig}, \citenamefont {Mock},\ and\ \citenamefont
  {Smith}(2006)}]{Sch06}%
  \BibitemOpen
  \bibfield  {author} {\bibinfo {author} {\bibfnamefont {D.}~\bibnamefont
  {Schurig}}, \bibinfo {author} {\bibfnamefont {J.~J.}\ \bibnamefont {Mock}}, \
  and\ \bibinfo {author} {\bibfnamefont {D.~R.}\ \bibnamefont {Smith}},\ }\href
  {\doibase 10.1063/1.2166681} {\bibfield  {journal} {\bibinfo  {journal}
  {Applied Physics Letters}\ }\textbf {\bibinfo {volume} {88}},\ \bibinfo
  {pages} {041109} (\bibinfo {year} {2006})}\BibitemShut {NoStop}%
\bibitem [{\citenamefont {Padilla}\ \emph {et~al.}(2007)\citenamefont
  {Padilla}, \citenamefont {Aronsson}, \citenamefont {Highstrete},
  \citenamefont {Lee}, \citenamefont {Taylor},\ and\ \citenamefont
  {Averitt}}]{Pad07}%
  \BibitemOpen
  \bibfield  {author} {\bibinfo {author} {\bibfnamefont {W.~J.}\ \bibnamefont
  {Padilla}}, \bibinfo {author} {\bibfnamefont {M.~T.}\ \bibnamefont
  {Aronsson}}, \bibinfo {author} {\bibfnamefont {C.}~\bibnamefont
  {Highstrete}}, \bibinfo {author} {\bibfnamefont {M.}~\bibnamefont {Lee}},
  \bibinfo {author} {\bibfnamefont {A.~J.}\ \bibnamefont {Taylor}}, \ and\
  \bibinfo {author} {\bibfnamefont {R.~D.}\ \bibnamefont {Averitt}},\ }\href
  {\doibase 10.1103/PhysRevB.75.041102} {\bibfield  {journal} {\bibinfo
  {journal} {Physical Review B}\ }\textbf {\bibinfo {volume} {75}},\ \bibinfo
  {eid} {041102} (\bibinfo {year} {2007})}\BibitemShut {NoStop}%
\bibitem [{\citenamefont {Gupta}\ \emph {et~al.}(1996)\citenamefont {Gupta},
  \citenamefont {Garg}, \citenamefont {Bahl},\ and\ \citenamefont
  {Bhartia}}]{Gup96}%
  \BibitemOpen
  \bibfield  {author} {\bibinfo {author} {\bibfnamefont {K.~C.}\ \bibnamefont
  {Gupta}}, \bibinfo {author} {\bibfnamefont {R.}~\bibnamefont {Garg}},
  \bibinfo {author} {\bibfnamefont {I.}~\bibnamefont {Bahl}}, \ and\ \bibinfo
  {author} {\bibfnamefont {P.}~\bibnamefont {Bhartia}},\ }\href@noop {} {\emph
  {\bibinfo {title} {Microstrip Lines and Slotlines}}},\ \bibinfo {edition}
  {2nd}\ ed.\ (\bibinfo  {publisher} {Artech House},\ \bibinfo {year}
  {1996})\BibitemShut {NoStop}%
\bibitem [{\citenamefont {Caloz}\ and\ \citenamefont {Itoh}(2004)}]{Cal04}%
  \BibitemOpen
  \bibfield  {author} {\bibinfo {author} {\bibfnamefont {C.}~\bibnamefont
  {Caloz}}\ and\ \bibinfo {author} {\bibfnamefont {T.}~\bibnamefont {Itoh}},\
  }\href@noop {} {\bibfield  {journal} {\bibinfo  {journal} {IEEE Transaction
  on Antennas and Propagation}\ }\textbf {\bibinfo {volume} {52}},\ \bibinfo
  {pages} {1159} (\bibinfo {year} {2004})}\BibitemShut {NoStop}%
\bibitem [{\citenamefont {Sanada}, \citenamefont {Caloz},\ and\ \citenamefont
  {Itoh}(2004)}]{San04}%
  \BibitemOpen
  \bibfield  {author} {\bibinfo {author} {\bibfnamefont {A.}~\bibnamefont
  {Sanada}}, \bibinfo {author} {\bibfnamefont {C.}~\bibnamefont {Caloz}}, \
  and\ \bibinfo {author} {\bibfnamefont {T.}~\bibnamefont {Itoh}},\ }\href@noop
  {} {\bibfield  {journal} {\bibinfo  {journal} {IEEE Microwave and Wireless
  Component Letters}\ }\textbf {\bibinfo {volume} {14}},\ \bibinfo {pages} {68}
  (\bibinfo {year} {2004})}\BibitemShut {NoStop}%
\bibitem [{\citenamefont {Houzet}, \citenamefont {M\'elique},\ and\
  \citenamefont {Lippens}(2010)}]{Hou10}%
  \BibitemOpen
  \bibfield  {author} {\bibinfo {author} {\bibfnamefont {G.}~\bibnamefont
  {Houzet}}, \bibinfo {author} {\bibfnamefont {X.}~\bibnamefont {M\'elique}}, \
  and\ \bibinfo {author} {\bibfnamefont {D.}~\bibnamefont {Lippens}},\
  }\href@noop {} {\bibfield  {journal} {\bibinfo  {journal} {Progress in
  Electromagnetics Research}\ }\textbf {\bibinfo {volume} {12}},\ \bibinfo
  {pages} {225} (\bibinfo {year} {2010})}\BibitemShut {NoStop}%
\bibitem [{\citenamefont {Bahl}(2003)}]{Bah03}%
  \BibitemOpen
  \bibfield  {author} {\bibinfo {author} {\bibfnamefont {I.~J.}\ \bibnamefont
  {Bahl}},\ }\href@noop {} {\emph {\bibinfo {title} {{Lumped Element for RF and
  Microwave Circuits}}}}\ (\bibinfo  {publisher} {Artech House},\ \bibinfo
  {year} {2003})\BibitemShut {NoStop}%
\bibitem [{\citenamefont {Alley}(1970)}]{All70}%
  \BibitemOpen
  \bibfield  {author} {\bibinfo {author} {\bibfnamefont {G.~D.}\ \bibnamefont
  {Alley}},\ }\href@noop {} {\bibfield  {journal} {\bibinfo  {journal} {IEEE
  Transactions on Microwave Theory and Techniques}\ }\textbf {\bibinfo {volume}
  {MTT-18}},\ \bibinfo {pages} {1028} (\bibinfo {year} {1970})}\BibitemShut
  {NoStop}%
\bibitem [{\citenamefont {Chen}\ \emph {et~al.}(2004)\citenamefont {Chen},
  \citenamefont {Grzegorczyk}, \citenamefont {Wu}, \citenamefont {J.~Pacheco},\
  and\ \citenamefont {Kong}}]{Che04}%
  \BibitemOpen
  \bibfield  {author} {\bibinfo {author} {\bibfnamefont {X.}~\bibnamefont
  {Chen}}, \bibinfo {author} {\bibfnamefont {T.~M.}\ \bibnamefont
  {Grzegorczyk}}, \bibinfo {author} {\bibfnamefont {B.-I.}\ \bibnamefont {Wu}},
  \bibinfo {author} {\bibfnamefont {J.}~\bibnamefont {J.~Pacheco}}, \ and\
  \bibinfo {author} {\bibfnamefont {J.~A.}\ \bibnamefont {Kong}},\ }\href@noop
  {} {\bibfield  {journal} {\bibinfo  {journal} {Physical Review E}\ }\textbf
  {\bibinfo {volume} {70}},\ \bibinfo {pages} {016608} (\bibinfo {year}
  {2004})}\BibitemShut {NoStop}%
\bibitem [{\citenamefont {Smith}\ \emph {et~al.}(2005)\citenamefont {Smith},
  \citenamefont {Vier}, \citenamefont {Koschny},\ and\ \citenamefont
  {Soukoulis}}]{Smi05}%
  \BibitemOpen
  \bibfield  {author} {\bibinfo {author} {\bibfnamefont {D.~R.}\ \bibnamefont
  {Smith}}, \bibinfo {author} {\bibfnamefont {D.~C.}\ \bibnamefont {Vier}},
  \bibinfo {author} {\bibfnamefont {T.}~\bibnamefont {Koschny}}, \ and\
  \bibinfo {author} {\bibfnamefont {C.~M.}\ \bibnamefont {Soukoulis}},\
  }\href@noop {} {\bibfield  {journal} {\bibinfo  {journal} {Physical Review
  E}\ }\textbf {\bibinfo {volume} {71}},\ \bibinfo {pages} {036617} (\bibinfo
  {year} {2005})}\BibitemShut {NoStop}%
\bibitem [{\citenamefont {Withayachumnankul}\ and\ \citenamefont
  {Abbott}(2009)}]{Wit09}%
  \BibitemOpen
  \bibfield  {author} {\bibinfo {author} {\bibfnamefont {W.}~\bibnamefont
  {Withayachumnankul}}\ and\ \bibinfo {author} {\bibfnamefont {D.}~\bibnamefont
  {Abbott}},\ }\href@noop {} {\bibfield  {journal} {\bibinfo  {journal} {IEEE
  Photonics Journal}\ }\textbf {\bibinfo {volume} {1}},\ \bibinfo {pages} {99}
  (\bibinfo {year} {2009})}\BibitemShut {NoStop}%
\end{thebibliography}

%

\end{document}